\documentclass[aps,pra,twocolumn,preprintnumbers,showpacs,superscriptaddress,amsmath,amssymb]{revtex4}
\usepackage{dcolumn}
\usepackage{bm}
\usepackage{amssymb}
\usepackage[german, english]{babel}
\usepackage{graphicx}
\usepackage{color}
\usepackage[letterpaper,total={7in,9.5in},top=0.75in,left=0.75in]{geometry}

\begin{document}
\title{Double-degenerate Bose-Fermi mixture of strontium}
\author{{Meng Khoon} Tey}
\affiliation{Institut f\"ur Quantenoptik und Quanteninformation (IQOQI),
\"Osterreichische Akademie der Wissenschaften, 6020 Innsbruck,
Austria}
\author{Simon Stellmer}
 \affiliation{Institut f\"ur Quantenoptik und Quanteninformation (IQOQI),
\"Osterreichische Akademie der Wissenschaften, 6020 Innsbruck,
Austria}
\affiliation{Institut f\"ur Experimentalphysik und
Zentrum f\"ur Quantenphysik, Universit\"at Innsbruck,
6020 Innsbruck, Austria}
\author{Rudolf Grimm}
 \affiliation{Institut f\"ur Quantenoptik und Quanteninformation (IQOQI),
\"Osterreichische Akademie der Wissenschaften, 6020 Innsbruck,
Austria}
\affiliation{Institut f\"ur Experimentalphysik und
Zentrum f\"ur Quantenphysik, Universit\"at Innsbruck,
6020 Innsbruck, Austria}
\author{Florian Schreck}
\affiliation{Institut f\"ur Quantenoptik und Quanteninformation (IQOQI),
\"Osterreichische Akademie der Wissenschaften, 6020 Innsbruck,
Austria}

\date{\today}


\begin{abstract}
We report on the attainment of a spin-polarized Fermi sea of $^{87}$Sr in thermal contact with a Bose-Einstein condensate (BEC) of $^{84}$Sr. Interisotope collisions thermalize the fermions with the bosons during evaporative cooling. A degeneracy of $T/T_F=0.30(5)$ is reached with $2\times 10^4$ $^{87}$Sr atoms together with an almost pure $^{84}$Sr BEC of $10^5$ atoms.
\end{abstract}

\maketitle

Ground-breaking experiments with ultracold Fermi gases \cite{Inguscio2006ufg, Giorgini2008tou} have opened possibilities to study fascinating phenomena, as the BEC-BCS crossover, with a high degree of control. Most experiments have been performed with the two alkali fermions $^{40}$K and $^6$Li. Fermions with two valence electrons, like $^{43}$Ca, $^{87}$Sr, $^{171}$Yb, and $^{173}$Yb, have a much richer internal state structure, which is at the heart of recent proposals for quantum computation and simulation \cite{Gorshkov2010tos,Cazalilla2009ugo,FossFeig2009ptk,Hermele2009mio,Gerbier2009gff,Daley2008qcw,Gorshkov2009aem}. Unlike bosonic isotopes of these elements, the fermions have a nuclear spin, which decouples from the electronic state in the $^1S_0$ ground state. This gives rise to a $SU(N)$ spin symmetry, where $N$ is the number of nuclear spin states, which is 10 for $^{87}$Sr. This symmetry can lead to new quantum phases in optical lattices \cite{Gorshkov2010tos,Cazalilla2009ugo,FossFeig2009ptk,Hermele2009mio}, like the chiral spin liquid. Non-Abelian gauge potentials can be realized by engineering state dependent lattices \cite{Gerbier2009gff}. In addition, the nuclear spin can be used to robustly store quantum information, which can be manipulated using the electronic structure \cite{Daley2008qcw,Gorshkov2009aem}. Double-degenerate Bose-Fermi mixtures extend the possibilities even further, allowing to study phase-separation and the effects of mediated interactions.

Evaporative cooling of ultracold atoms to quantum degeneracy relies on elastic collisions to thermalize the sample. Identical fermions do not collide at low temperatures, therefore mixtures of spins \cite{Demarco1999oof,Granade2002aop,Fukuhara2007dfg,DeSalvo2010dfg,Taie2010ros}, isotopes \cite{Truscott2001oof,Schreck2001qbe,Mcnamara2006dbf}, or elements \cite{Roati2002fbq,Hadzibabic2002tsm,Silber2005qdm} are used for evaporation. $^{87}$Sr has a large nuclear spin of $I=9/2$, which leads to a tenfold degenerate ground state. Recently, $^{87}$Sr in a mixture of these states was cooled to quantum degeneracy \cite{DeSalvo2010dfg}. In this rapid communication, we report on the attainment of a spin-polarized quantum degenerate Fermi gas of $^{87}$Sr together with a BEC of $^{84}$Sr. Interisotope collisions are used to thermalize the fermions with the bosons. The system provides a clean starting point for the exploration of alkaline-earth Bose-Fermi mixtures.


\begin{figure}
\includegraphics[width=\columnwidth]{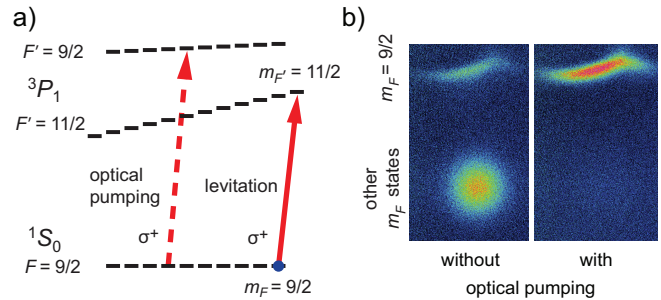}
\caption{\label{fig:Fig1} (Color online) Optical pumping and state detection of $^{87}$Sr. a) Internal states and transitions of $^{87}$Sr used for optical pumping and detection of spin polarization. The optical pumping beam (dashed arrow) is scanned in frequency to address all $m_F$ states. Internal state detection is performed by levitating only the $m_F=9/2$-state atoms against gravity using a $\sigma^+$-polarized laser beam on a cycling transition (solid arrow). b) Levitated $m_F=9/2$-state atoms and atoms in all other $m_F$ states, 15\,ms after switching off the dipole trap.
}
\end{figure}


Our scheme to generate a BEC of $^{84}$Sr is described in Ref. \cite{Stellmer2009bec}. We have extended this scheme to the preparation of Bose-Fermi mixtures. In the following, we briefly summarize the basic preparation steps, while technical details will be presented elsewhere \cite{StellmerDetails}. We take advantage of atoms in the metastable $5s5p\,^3P_2$ state that can be trapped in the quadrupole magnetic field of the magneto-optical trap (MOT). These atoms are automatically produced when operating a ``blue MOT'' on the $5s^{2}\,{^1S_0}-5s5p\,{^1P_1}$ transition at a wavelength of 461\,nm. The 461-nm laser system is initially tuned to the $^{84}$Sr line and metastable-state bosons are accumulated in the magnetic trap. After 5\,s, the laser frequency is shifted by 210\,MHz to the $^{87}$Sr line and metastable-state fermions are added to the magnetic trap for another 6\,s \cite{Poli2005cat}.


Further cooling and density increase of the mixture is achieved by operating ``red MOTs'' for each isotope simultaneously on the 7.4\,kHz linewidth $^1S_0-{^3P_1}$ intercombination lines at 689\,nm. Since $^{87}$Sr has a nuclear spin, its $^3P_1$ state is split into three hyperfine states with $F'=\,$11/2, 9/2, and 7/2. The $^{87}$Sr MOT uses the transitions to the $F'=9/2$ and $F'=11/2$ states simultaneously \cite{Mukaiyama2003rlr}. To increase the capture velocity of the red MOTs, we frequency modulate the light, producing sidebands, which cover a detuning range from a few ten kHz to a few MHz to the red of the transition. To load the red MOTs, the metastable-state atoms in the magnetic trap are optically pumped to the $^1S_0$ ground state using the $5s5p\,{^3P_2}-5s5d\,{^3D_2}$ transition at 497\,nm. The nuclear spin of $^{87}$Sr leads to a splitting of the $^3P_2$ and $^3D_2$ states into five hyperfine states each. We use two of the 13 transitions between these states for repumping \cite{StellmerDetails}. After loading, the MOT is compressed by reducing the MOT beam intensity and ramping off the frequency modulation, resulting in a colder and denser sample. After compression, the $^{84}$Sr and $^{87}$Sr MOTs are spatially separated as a consequence of different red MOT dynamics \cite{StellmerDetails, Katori1999mot, Mukaiyama2003rlr}. The MOTs each contain $\sim10^7$ atoms at a temperature of $\sim2\,\mu\textrm{K}$.


For evaporative cooling, the atoms are transferred into a crossed-beam optical dipole trap (ODT) based on a broadband ytterbium fiber laser operating at 1075\,nm. The trapping geometry consists of a horizontal and a nearly vertical beam with waists of 80\,$\mu$m. During the red MOT compression phase, the horizontal and vertical beams are ramped to a power of 4.2\,W and 1.6\,W, respectively, corresponding to an average trap oscillation frequency of 170\,Hz and a trap depth of 13\,$\mu$K.

Loading of the dipole trap is achieved in two sequential stages to overcome the spatial separation of the MOTs, first $^{87}$Sr then $^{84}$Sr. For each stage, the center of the respective MOT is overlapped with the cross of the ODT by shifting the center of the quadrupole field. For optimum loading, we adjust the intensities and detuning of the MOT beams before switching off the MOT beams. After loading both isotopes, the ODT power is suddenly increased to 6\,W in the horizontal and 2\,W in the vertical beam, resulting in a trap depth of 22\,$\mu$K and an average trap frequency of 200\,Hz.


In order to obtain a spin-polarized sample of $^{87}$Sr, we perform optical pumping on the $^1S_0 (F=9/2) - {^3P_1} ( F'=9/2)$ transition with a $\sigma^+$-polarized laser beam \cite{OpticalPumpingDetails} parallel to a homogeneous guiding field of 3\,G; see Fig.~\ref{fig:Fig1} (a). The $m_F=9/2$ state is a dark state of this transition. The frequency of the optical pumping light is scanned over the transitions of the different $m_F$ states, which are spread over 2\,MHz. After 30\,ms of optical pumping, the temperature of the mixture has increased to 3.4\,$\mu$K.


To perform state detection, we selectively levitate only the $m_F=9/2$-state atoms to separate them from atoms in all other states after switching off the ODT. An upward propagating $\sigma^+$-polarized laser beam near the $^1S_0 (F=m_F=9/2) - {^3P_1}(F'=m_{F'}=11/2)$ cycling transition is used to levitate the atoms. The intensity and detuning of this beam are adjusted to optimize the levitation. At the magnetic field of 3\,G, the splitting between adjacent $m_{F'}$ states is $\sim1.1\,\textrm{MHz}$, which is about 10 times the power-broadened linewidth. Atoms in the $m_F\neq9/2$ states are therefore not levitated. Figure~\ref{fig:Fig1} (b) shows the atoms in the $m_F=9/2$ state separated from the other atoms after time of flight, with and without optical pumping. From these images we conclude that more than 95\% of the atoms are in the same spin state.

\begin{figure}
\includegraphics[width=\columnwidth]{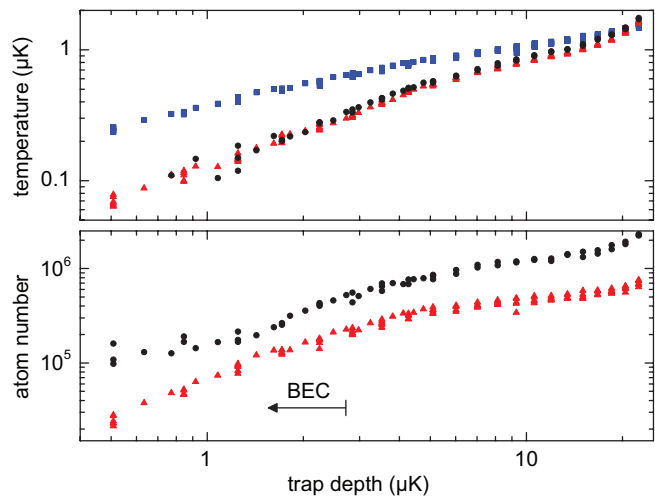}
\caption{\label{fig:Fig2} (Color online) Temperature and atom number in dependence of the trap depth during evaporative cooling. The upper panel shows the temperature of $^{84}$Sr (black circles) and the temperature (red triangles) and Fermi temperature (blue squares) of $^{87}$Sr. The lower panel shows the atom number of $^{84}$Sr (black circles) and $^{87}$Sr (red triangles). A BEC forms below a trap depth of 2.8\,$\mu$K (arrow).}
\end{figure}


After 500\,ms plain evaporation, the mixture thermalizes at $1.7\,\mu$K with $2.2\times 10^6$ $^{84}$Sr atoms at a phase-space density of 0.3 and $7\times 10^5$ $^{87}$Sr atoms at $T/T_F=1.2$, already close to quantum degeneracy. At this point, the elastic collision rate of bosons with fermions is ${\sim170\,\textrm{s}^{-1}}$. The collision rate between bosons of $\sim1600\,\textrm{s}^{-1}$ is much higher because of the larger scattering length ($a_{84,84}=123\,a_0$ versus $a_{84,87}=-56\,a_0$ \cite{Stein2008fts,Martinezdeescobar2008tpp}) and because of the factor of two resulting from Bose enhancement. These conditions are well suited for evaporative cooling to quantum degeneracy. To force evaporation, we reduce the power of both dipole trap beams with 1/e time constants of 5.7\,s for the horizontal and 9.0\,s for the vertical beam, starting directly after optical pumping. To characterize the performance of evaporative cooling and ultimately detect quantum degeneracy, we stop forced evaporation after a varying time between 0 and 8\,s, wait 500\,ms to ensure thermalization of the sample, and take absorption images after a time of flight $t_{\rm TOF}$ of 10\,ms \cite{absImageDetails}. This expansion time is sufficiently large for the observed density distribution to correspond to the in-situ momentum distribution, even for the lowest trap frequencies of 85\,Hz examined. We determine the atom number and temperature using Bose-Einstein or Fermi-Dirac distribution fits. The results show that both bosons and fermions are evaporated and are well thermalized with each other (Fig.~\ref{fig:Fig2}). To demonstrate the importance of interisotope collisions for the cooling of spin-polarized $^{87}$Sr, we perform the same evaporative cooling sequence in absence of $^{84}$Sr. The fermions are lost faster than in presence of bosons and do not reach as low temperatures. In addition, the momentum distributions are in this case inconsistent with Fermi-Dirac distributions, which is expected since the sample cannot thermalize. During the evaporation of both isotopes, the BEC phase transition is detected after 6\,s by the appearance of a bimodal distribution \cite{Stellmer2009bec,deEscobar2009bec}. At that moment $5\times 10^5$ bosons remain at a temperature of 300\,nK.

\begin{figure}
\includegraphics[width=\columnwidth]{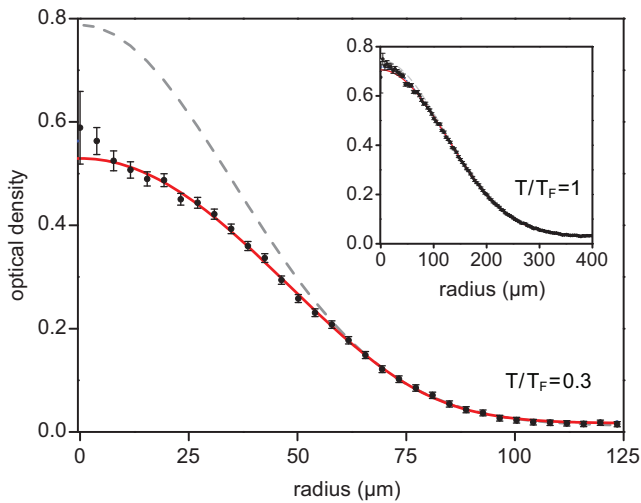}
\caption{\label{fig:fig3} (Color online)
Azimuthally averaged density distributions of a degenerate Fermi gas at $T/T_F=0.3$ and a thermal fermionic cloud at $T/T_F=1$ (inset) after 10\,ms of free expansion. Fermi-Dirac distributions (solid red lines) fit both measurements well. The quantum degenerate sample deviates in shape from a Gaussian (dotted blue line). The difference between the degenerate and thermal samples becomes even more evident when fitting a Gaussian to only the outer wings, outside the disk with radius $\sqrt{2}w$ (dashed line), where $w$ is the 1/e width of a Gaussian fit to the full distribution. For $T/T_F=1$ (inset), the difference between the three fits is barely visible. }
\end{figure}


Fermionic degeneracy manifests itself in subtle changes of the momentum distribution. Pauli blocking limits the occupation of low momentum states to one fermion per state, leading to a flatter and wider momentum distribution than expected classically. This change in shape is evident in the azimuthally averaged time-of-flight density distributions shown in Fig.~\ref{fig:fig3}. The inset shows a thermal cloud at the beginning of evaporation and the main figure a colder sample after 7.6\,s of evaporation. The momentum distribution after evaporation is very well described by a Fermi-Dirac distribution, but not by Gaussian fits.

\begin{figure}
\includegraphics[width=\columnwidth]{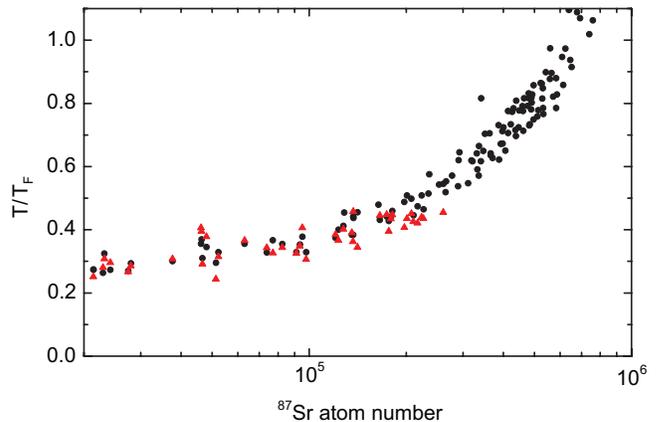}
\caption{\label{fig:Fig4} (Color online) $T/T_F$ in dependence of the $^{87}$Sr atom number during evaporative cooling. $T/T_F$ is determined from Fermi fits to absorption images. Two independent fit parameters are used to calculate $T/T_F$, the width of the fitted function (filled circles) and the fugacity (triangles). The statistical error visible in the scatter of the data is larger than systematic uncertainties.
}
\end{figure}

To quantify the degeneracy, we determine the ratio of the temperature $T$ and the Fermi temperature $T_F$ from fits to absorption images. For a harmonic potential and a time of flight larger than the trap oscillation periods, the spatially once integrated Fermi-Dirac density distribution can be closely approximated by \cite{DeMarcoPhD}
\begin{equation}
n(\rho)=A\, \mathrm{Li}_2\left(-\zeta e^{-\frac{\rho^2}{2 \sigma^2}} \right),
\end{equation}
where $\mathrm{Li}_n$ is the $n^{\textrm{th}}$-order polylogarithm, $\rho$ the radial displacement from the center of the atom cloud, $\sigma$ the width of the cloud, and $\zeta$ the fugacity. $A$, $\zeta$ and $\sigma$ are fit parameters. We use two independent methods to determine $T/T_F$. The first one rests upon the size of the cloud $\sigma$, which is related to the temperature by $T=m\sigma^2/k_B t_{\rm TOF}^2$. $T_F$ is calculated from the atom number $N$ and the average trap frequency $\overline{f}$ using $T_F=(6N)^{1/3}\hbar 2\pi\overline{f}/k_B$. The second method is based on the examination of the shape of the distribution, characterized by the fugacity $\zeta$. $T/T_F$ is directly related to the fugacity through the relation $T/T_F=[-6 \mathrm{Li}_3(-\zeta)]^{-1/3}$. The second method is only applied for fugacities greater than 2, corresponding to $T/T_F\leq0.46$. Using numerical simulations, we verified that mean field effects between the BEC and the degenerate Fermi gas, as well as the increased anharmonicities of the trapping potential in the vertical direction due to gravitational sagging, do not influence the determination of $T/T_F$ under our conditions. Figure~\ref{fig:Fig4} shows $T/T_F$ in dependence of the $^{87}$Sr atom number using the same data set as for Fig.~\ref{fig:Fig2}. Both methods to determine $T/T_F$ agree well. With $10^5$ $^{87}$Sr atoms left we obtain $T/T_F=0.35(5)$ with $T=160(15)$\,nK and $T_F=450(20)\,$nK. The lowest degeneracy, $T/T_F=0.30(5)$, is reached with $2\times 10^4$ fermions at $T=65(5)\,$nK. The efficiency of evaporation is reduced in the quantum degenerate regime because of Pauli blocking \cite{Demarco2001pbo} and the superfluidity of the BEC \cite{Timmermans1998sis,Chikkatur2000sae}. This reduction, together with heating and loss, limits the obtainable fermionic degeneracy.

For the deepest evaporation examined, the potential depth corresponds only to two times the Fermi energy. The shallow potential can lead to increased evaporation of fermions in the $m_F=9/2$ state compared to bosons and fermions in other spin states. This preferential evaporation can reduce the polarization of the fermionic sample. We verified that more than 90\% of the fermions occupy the $m_F=9/2$ state even at the end of evaporation.

\begin{figure}
\includegraphics[width=\columnwidth]{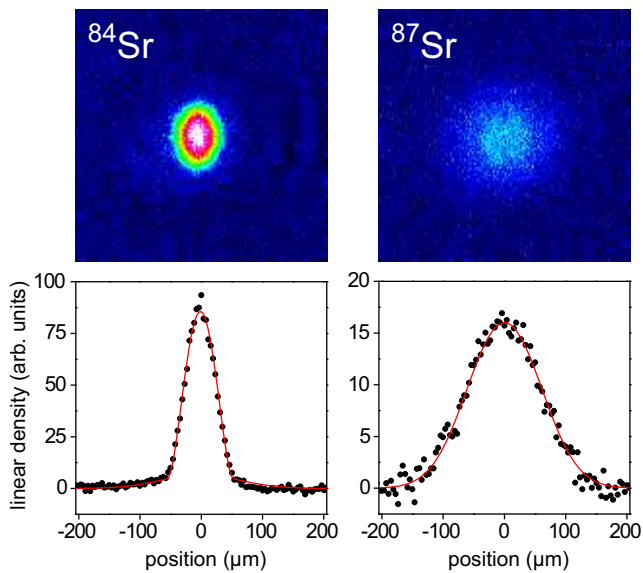}
\caption{\label{fig:fig5} (Color online) Time-of-flight absorption images of a nearly pure $^{84}$Sr BEC and a $^{87}$Sr Fermi gas at $T/T_F=0.45$ after 15\,ms of expansion.
}
\end{figure}

Our degenerate Fermi sea is accompanied by a BEC containing $10^5$ atoms. Figure~\ref{fig:fig5} shows time-of-flight absorption images of a BEC and a Fermi gas with $T/T_F=0.45$, taken 15\,ms after release from a cigar-shaped trap with trap frequency ratio of 1.5. The BEC expansion is mean-field driven and leads to an inverted aspect ratio. By contrast, the $^{87}$Sr image shows the isotropic momentum distribution of the fermions.

In conclusion, we have created a double-degenerate Bose-Fermi mixture with the fermions prepared in a single spin state. In the future, we plan to implement optical pumping and state detection methods for the controlled preparation of fermions in multiple internal states. Loading the mixture into an optical lattice and further cooling of the sample are the next steps towards the exploration of the $SU(10)$ symmetric Hubbard model, and quantum computation and simulation with strontium.

We thank Andrew Daley and Sebastian Diehl for stimulating discussions on the prospects of strontium Bose-Fermi mixtures for future experiments.


\begin{thebibliography}{34}
\bibitem[{Opt()}]{OpticalPumpingDetails}
\bibinfo{note}{The laser beams used for optical pumping and state detection are
  identical to the red MOT beams oriented vertically upwards and have an
  intensity of $10\,\mu$W/cm$^2$ and $150\,\mu$W/cm$^2$, respectively. The
  other five MOT beams are blocked by shutters.}
\bibitem[{abs()}]{absImageDetails}
\bibinfo{note}{The absorption images are taken using a cycling transition.
  Images of $^{84}$Sr and $^{87}$Sr are taken in consecutive runs of the
  experiment.}
\end{thebibliography}
\end{document}